\documentclass[12pt]{article}
\usepackage{amsmath,amssymb,amsthm,latexsym}
\usepackage{stmaryrd,wasysym,upgreek,mathrsfs,dsfont}
\usepackage[english]{babel}
\usepackage{graphicx,color}
\usepackage{amscd}

\usepackage[pdftex]{hyperref}
\newtheorem{lemma}{Lemma}[section]
\newtheorem{example}{Example}[section]
\newtheorem{conjecture}{Conjecture}[section]

\newtheorem{definition}{Definition}[section]
\newtheorem{theorem}{Theorem}[section]
\newcommand{\be}{\begin{equation}}
\newcommand{\ee}{\end{equation}}
\newcommand{\bea}{\begin{eqnarray}}
\newcommand{\eea}{\end{eqnarray}}

\newcommand{\tr}{\mbox{tr}}

\newcommand{\cA}{{\cal{A}}}

\newcommand{\cT}{{\cal{T}}}

\newcommand{\cF}{{\cal{F}}}

\begin{document}

 \title{How are Feynman graphs resummed by the Loop Vertex Expansion?}
\author{Vincent Rivasseau,\  Zhituo Wang\\
Laboratoire de Physique Th\'eorique, CNRS UMR 8627,\\
Universit\'e Paris XI,  F-91405 Orsay Cedex, France\\
E-mail: rivass@th.u-psud.fr,  zhituo.wang@th.u-psud.fr}

\maketitle
\begin{abstract}
The purpose of this short letter is to clarify which 
set of pieces of Feynman graphs are resummed in
a Loop Vertex Expansion, and to formulate a conjecture on the 
$\phi^4$ theory in non-integer dimension.
\end{abstract}

\begin{flushright}
LPT-20XX-xx
\end{flushright}
\medskip


\medskip

\section{Introduction}

In quantum field theory  (hereafter QFT) any connected (i.e. interesting) quantity is written as
a sum of amplitudes for a certain category of connected graphs
\be  S = \sum_{G\;\; {\rm connected}}  \cA_G  \label{ordinar}
\ee
but this formula is not a valid definition of $S$ since
usually 
\be  \sum_{G\;\; {\rm connected}} \vert  \cA_G \vert = \infty .
\ee
This phenomenon, known since \cite{Dys},
is basically due to the very large number of elements at order $n$ in the species \cite{BLL} 
of Feynman graphs. Accordingly the {\it generating functional for the Feynman graphs species},
namely the series  $\sum_n  \frac{\lambda^n a_n}{n!}$, where $a_n$ is the number of Feynman graphs at order $n$, has zero radius of convergence as power series in $\lambda$.
We call such a species a {\it  proliferating species}. In zero space-time dimension, quantum field theory 
reduces to this generating functional, hence to graphs counting. In higher dimensions 
quantum field theory is in fact a {\it weighted} such
species, that is Feynman graphs have to be pondered with weights,  called Feynman amplitudes. For an introduction to the structure of Feynman graphs, see \cite{tutte}.
Nevertheless these Feynman 
amplitudes tend to behave as $K^n $ at order $n$ (at least in low dimensions). Hence the 
perturbation series eg for the $\phi^4$ Euclidean Bosonic quantum field theory
tends to behave as $ \sum_n  (-\lambda)^n K^n n! $ and it has
been proved to have zero radius of convergence in one, two and three 
dimensions (\cite{Jaf,CR}). Nothing is yet known for sure in dimension 4 but there are 
strong reasons to expect also the renormalized Feynman series to diverge there as well (see \cite{Riv} and references therein).

In contrast Cayley's theorem, which states that the total number of labeled trees at order $n$ is 
$n^{n-2}$, implies that the species of {\it trees} is {\it not} proliferating. This fact can be related to the
local existence theorems for flows in classical mechanics, since classical perturbation theory
is indexed by trees \cite{LinPoinc}. These theorems have no quantum counterpart, but
constructive theory can be seen as various recipes to replace the ordinary 
divergent Feynman graph expansions
by convergent ones, indexed by trees rather than graphs \cite{Rivasseau:2000gd}. It can
therefore be considered a  bridge between QFT and classical mechanics, since it repacks the loops which are the fundamental feature of QFT, and brings the expansion closer to those of classical mechanics. Historically constructive theory used cumbersome non canonical tools
borrowed from lattice statistical mechanics, such as cluster expansions which did not respect the rotational invariance of the underlying theory \cite{GJ,Riv}.
The Loop Vertex Expansion  \cite{R1,MR1} is a more canonical way to replace the ordinary perturbative divergent expansion by a convergent one, which in principle allows to 
compute quantities to arbitrary accuracy.

One of us (VR) was recently asked exactly which (pieces of) Feynman graphs
are resummed by this expansion. The answer is contained
in the initial papers, but perhaps not easy to extract. 
The purpose of this little note is therefore to explain more explicitly exactly
which pieces of which Feynman graphs of different orders are combined together by the loop vertex expansion to create a convergent expansion. This reshuffling is fully explicited up to third order
for the simplest of all possible examples,  namely the $\phi^4_0$ quantum field theory. 
Finally we also propose a conjecture, which, if true, would allow to define QFT in non-integer dimensions
of space-time.

\section{Relative Tree Weights in a Graph}

A graph may contain many (spanning) forests, and a forest can be extended into
many graphs with loops. So the relationship between graphs and their spanning forests
is not trivial.

The forest formula which we use \cite{BK,AR1}
can be viewed as a tool to associate to any 
pair made of a graph $G$ and a spanning forest $\cF \subset G$ 
a unique rational number or weight $w(G,\cF)$ between 0 and 1, called 
the relative weight of $\cF$ in $G$. 

The numbers $w(G,\cF)$ are multiplicative over disjoint unions \footnote{And 
also over vertex joints of graphs, just as in the universality theorem for the Tutte polynomial.}.
Hence it is enough to give the formula for $(G,\cF)$ only when $G$ is {\it connected}
and $\cF=\cT$ is a spanning {tree} in it\footnote{It is enough in fact to compute such weights 
for 1-particle irreducible and 1-vertex-irreducible graphs, then multiply them 
in the appropriate way for the general case.}. 

The definition of these weights is
\begin{definition}
\be  w(G,\cT) =  \int_0^1 \prod_{\ell \in \cT} dw_\ell   
\prod_{\ell \not\in \cT} x^\cT_{\ell}(\{w\})
\ee
where $x^\cT_{\ell}(\{w\})$ is the infimum over the $w_{\ell'}$ parameters 
over the lines $\ell'$ forming the {\it unique} path in $\cT$ joining the ends of $\ell$.
\end{definition}
\begin{lemma} The relation
\be  \sum_{\cF\subset G}  w(G,\cF) =1  \label{bary}
\ee
holds for any connected graph $G$.
\end{lemma}
\noindent{\bf Proof}  It is a simple consequence of the forest formula \cite{BK,AR1} applied to 
the lines of the graph $G$.

\subsection{Examples}

For a fixed spanning tree inside a graph, we call {\it loop lines} the lines not in the tree.

\begin{figure}[!htb]
\centering
\includegraphics[scale=0.8]{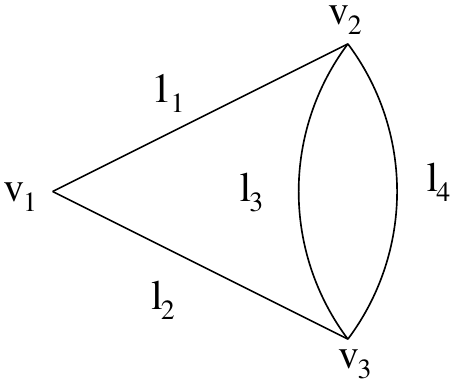}
\caption{A first example}
\label{exa}
\end{figure}

Consider the graph $G$ of Figure \ref{exa}.
There are $5$ spanning trees inside this graph:
$$
\cT_{12}=\{l_1, l_2\},\     \cT_{13}=\{l_1, l_3\},\ \cT_{14}=\{l_1, l_4\},\ \cT_{23}=\{l_2, l_3\},\ \cT_{24}=\{l_2, l_4\}.
$$
For example, for the tree $\cT_{12} =\{l_2, l_4\}$, the loop lines are $l_1$ and $l_3$.

To take into account the 
weakening factors $x^\cT_{\ell}(\{w\})$ of (\ref{bary}) for each loop line $\ell$, it is convenient to decompose the integration domain $[0,1]^{\vert \cT\vert}$  into $\vert \cT\vert !$ sectors corresponding to complete orderings of the $w_\ell$ parameters for $\ell \in \cT$.

Let us compute in this way the relative weights of the five trees of $G$. First consider the contribution of the tree $\cT_{12}$. In this case the loop lines are $l_3$ and $l_4$. For each of them we have a factor $\inf(w_1 w_2)$. Hence
\begin{eqnarray}
w(G, \cT_{12})&=&\int_0^1\int_{0}^1 dw_1d w_2 [\inf(w_1, w_2)]^2 \nonumber\\
&=&2 \int_0^1 dw_2\int_0^{w_2}dw_1 w_1^2=\frac{2}{12}=\frac{1}{6}.\nonumber
\end{eqnarray}
Next we consider the spanning tree $\cT_{13}$. In this case the ''loop lines`` are $l_2$ which connects the vertices $v_1$ and $v_3$ and $l_4$ which connects $v_2$ and $v_3$.  So we have:
 \begin{eqnarray}
w(G, \cT_{13})&=&\int_{w_1<w_3} dw_1\int d w_3 \inf(w_1 w_3) w_3\\&+&
\int_{w_3<w_1} dw_1\int d w_3 \inf(w_1 w_3) w_3\nonumber\\
&=&\int_0^1 dw_3\int_0^{w_3}dw_1 w_1 w_3+\int_0^1 dw_1\int_0^{w_1}dw_3 w_3^2=\frac{1}{8}+\frac{1}{12}=\frac{5}{24}.\nonumber
\end{eqnarray}
With the same method we find that 
\begin{equation}
w(G, \cT_{14})=w(G, \cT_{24})=w(G, \cT_{23})=\frac{5}{24},
\end{equation}
and we have
\begin{equation}
 \sum_{\cT\in G} w(G,\cT) =\frac{1}{6} +4. \frac{5}{24} =1 .
\end{equation}

\begin{figure}[!htb]
\centering
\includegraphics[scale=0.8]{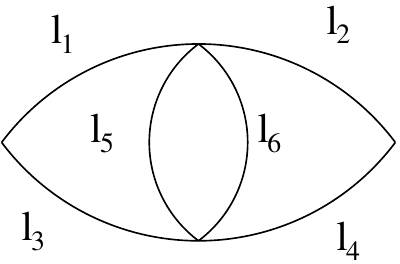}
\caption{Example 2-the eye graph}
\label{eyea}
\end{figure}

\begin{figure}[!htb]
\centering
\includegraphics[scale=0.8]{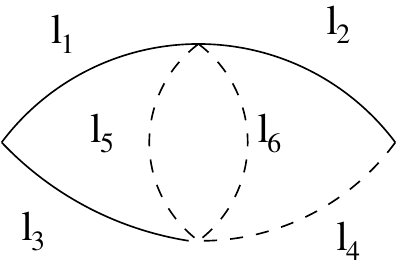}
\caption{The spanning tree $\{l_1, l_2, l_3\}$}
\label{eyeb}
\end{figure}

Let us treat a second example.
Consider the graph $G'$ of Fig.  \ref{eyea}, which has 6 edges:
\begin{equation}
 \{l_1, l_2, l_3, l_4, l_5, l_6\} .
\end{equation}
To each edge $l_i$ we associate a factor $w_i$.
There are 12 spanning trees:
\begin{eqnarray}
  \{l_1, l_2, l_3\},  \{l_1, l_2, l_4\},  \{l_1, l_3, l_4\},  \{l_2, l_3, l_4\},  \{l_1, l_2, l_5\},\{l_1, l_2, l_6\},\nonumber\\  \{l_3, l_4, l_5\}, 
\{l_3, l_4, l_6\},  \{l_1, l_5, l_4\},  \{l_1, l_6, l_4\},  \{l_3, l_5, l_2\},  \{l_3, l_6, l_2\}.
\end{eqnarray}
Let us compute the relative weight for each of these spanning trees in $G'$.
First of all consider $\cT_{123}=\{l_1, l_2, l_3\}$. The other edges are drawn in dotted lines. See figure(\ref{eyeb})
As is easily seen the corresponding loop lines are $l_4$, $l_5$ and $l_6$. The weakening
factor for $l_5$ and $l_6$ $\inf(w_1, w_3)$and 
the weakening factor for $l_4$ is $\inf(w_1, w_2, w_3)$. Therefore we have
\begin{eqnarray}
&& w(G',\cT_{123})=\int_{0<w_1<w_2<w_3<1}dw_1 dw_2 dw_3\inf(w_1, w_3)^2\inf(w_1, w_2, w_3)\nonumber\\&+&\rm{other}\  \rm{permutations}\  \rm{of}\  w_1, w_2, w_3\nonumber\\
&=&\int_{w_1<w_2<w_3}dw_1 dw_2 dw_3 w_1^3+\int_{w_2<w_3<w_1}dw_1 dw_2 dw_3 w_3^2 w_2\nonumber\\&+&\int_{w_3<w_1<w_2}dw_1 dw_2 dw_3 w_3^3+\int_{w_2<w_1<w_3}dw_1 dw_2 dw_3 w_1^2 w_2\nonumber\\&+&\int_{w_3<w_2<w_1}dw_1 dw_2 dw_3 w_3^3 
+\int_{w_1<w_3<w_2}dw_1 dw_2 dw_3 w_1^3.\nonumber
\end{eqnarray}
We compute only two of the integrals explicitly as others are obtained 
by changing the names of variables. 
\begin{equation}
\int_{w_1<w_2<w_3}dw_1 dw_2 dw_3\  w_1^3=\int_0^1 dw_3\int_0^{w_3} dw_2 \int_0^{w_2}dw_1 w_1^3=  \frac{1}{120},
\end{equation}
\begin{equation}
 \int_{w_1<w_2<w_3}dw_1 dw_2 dw_3\  w_3^2\  w_2=\frac{1}{60} .
\end{equation}
So we have 
\begin{equation}
w(G',\cT_{123})=\frac{1}{120}\times 4+\frac{1}{60}\times 2=\frac{1}{15}.
\end{equation}

The relative weights in $G'$ of the spanning trees $\cT_{124}$, $\cT_{134}$ and $\cT_{234}$ are the same.

\begin{figure}[!htb]
\centering
\includegraphics[scale=0.8]{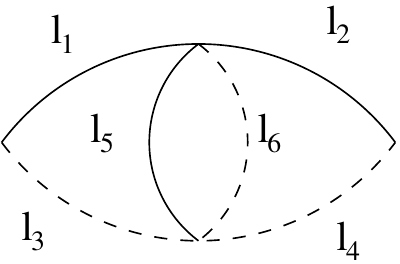}
\caption{The spanning tree $\{l_1, l_2, l_5\}$}
\label{eyec}
\end{figure}

Now we consider the tree $\{l_1, l_2, l_5\}$. (See figure \ref{eyec}). 
To the loop line $l_3 $ is associated a weakening factor $\inf(w_1, w_5)$. To the loop line $l_4$ is associated a weakening factor $\inf(w_2, w_5)$. To the loop line $l_6$
is associated a weakening factor  $w_5$. So we have
\begin{eqnarray}
&& w(G',\cT_{125})=\int_{w_1<w_2<w_5}dw_1 dw_2 dw_5\inf(w_1, w_5)\inf( w_2, w_5) w_5\\&+&\rm{other}\  \rm{permutations}\  \rm{of}\  w_1, w_2, w_5\nonumber\\
&=&\int_{w_1<w_2<w_5}dw_1 dw_2 dw_5 w_1 w_2 w_5+\int_{w_5<w_1<w_2}dw_1 dw_2 dw_5 w_5^3\nonumber\\&+&\int_{w_2<w_5<w_1}dw_1 dw_2 dw_5 w_5^2 w_2
+\int_{w_2<w_1<w_5}dw_1 dw_2 dw_5 w_1 w_2 w_5\nonumber\\&+&\int_{w_1<w_5<w_2}dw_1 dw_2 dw_5 w_5^2 w_1 +
\int_{w_5<w_2<w_1}dw_1 dw_2 dw_5 w_5^3.\nonumber
\end{eqnarray}
We have
\begin{equation}
 \int_{w_1<w_2<w_5}dw_1 dw_2 dw_5 w_1 w_2 w_5=\frac{1}{48},
\end{equation}
\begin{equation}
 \int_{w_5<w_1<w_2}dw_1 dw_2 dw_5 w_5^3=\frac{1}{120},
\end{equation}
\begin{equation}
 \int_{w_2<w_5<w_1}dw_1 dw_2 dw_5  w_2 w^2_5=\frac{1}{60}.
\end{equation}
Similarly we get
\begin{equation}
 w(G',\cT_{125})=\frac{1}{120}\times 2+\frac{1}{60}\times 2 +\frac{1}{48}\times 2=\frac{11}{120} .
\end{equation}

By the same method we find that this is also the relative weight of trees
$\cT_{126}, \cT_{345}, \cT_{346}, \cT_{125}, \cT_{145}, \cT_{146}, \cT_{235}$ and $\cT_{236}$.

We can check  again that
\begin{equation}
 \sum_{\cT \in G'} w(G', \cT) = 4.\frac{1}{15}+ 8.\frac{11}{120} =1.
\end{equation}

\section{Resumming Feynman Graphs}

\subsection{Naive Repacking}

Consider the expansion (\ref{ordinar}) of a connected quantity $S$.
The most naive way to reorder Feynman perturbation theory according to trees
rather than graphs is to insert for each graph the relation (\ref{bary}) 
\be  S = \sum_{G} A_G = \sum_G \sum_{\cT\subset G}  w(G,\cT) \cA_G 
\ee
and exchange the order of the sums over  $G$ and $\cT$. Hence it writes
\be  S =  \sum_\cT  \cA_\cT, \quad  \cA_\cT = \sum_{G \supset \cT}  w(G,\cT) \cA_G .
\ee

This rearranges the Feynman expansion according to trees, but each tree
has the same number of vertices as the initial graph. Hence it reshuffles
the various terms of a {\it given, fixed} order of perturbation theory. Remark
that if the initial graphs have say degree 4 at each vertex, only
trees with degree less than or equal to 4 occur in the rearranged
tree expansion.

For Fermionic theories this is typically sufficient and one has for small enough coupling
\be  \sum_{\cT}  \vert \cA_\cT \vert < \infty
\ee
because Fermionic graphs mostly compensate each other
at a fixed order by Pauli's principle; mathematically this is because these
graphs form a determinant and the size of a determinant is much less
than what its permutation expansion suggests. This is well known
\cite{Les,FMRT1,AR2}.

But this repacking fails for Bosonic theories, because the only compensations there
occur between graphs of different orders. Hence if we were to perform this naive reshuffling, eg on the 
$\phi^4_0$ theory we would still have
\be  \sum_{T}  \vert \cA_T \vert = \infty .
\ee

\section{The Loop Vertex Expansion}

The loop vertex expansion overcomes this difficulty by exchanging 
the role of vertices and propagators before applying the forest formula.
The corresponding regrouping is completely different and each tree resums
an infinite number of pieces of the previous graphs. It relies on 
a technical tool (which physicists call 
the intermediate field representation) 
which decomposes any interaction of degree higher than three
in terms of simpler three-body interactions.
It is particularly natural for 4-body interactions 
but can be generalized to higher interactions as well \cite{RW}.

This quite universal and powerful trick is linked to various deep physical 
and mathematical tools, such as the color 1/N expansion
and the Matthews-Salam and Hubbard-Stratonovich methods in physics
and the Kaufmann bracket of a knot and many similar ideas in mathematics.

It is easy to describe the intermediate field method in terms of functional integrals, as it is a simple generalization of the formula
\be  e^{- \lambda \phi^4/2}  = \int e^{- \sigma^2 /2} e^{i \sqrt \lambda \sigma \phi^2} d\sigma .
\ee
In this section we introduce the graphical procedure equivalent to this formula.

In the case of a $\phi^4$ graph $G$ each vertex has exactly four half-lines 
hence there are exactly three ways to pair these half-lines into two pairs. Hence
each fully labeled (vacuum) graph of order $n$ (with labels on vertices and half-lines),
which has $2n$ lines
can be decomposed exactly into $3^n$ labeled graphs $G'$
with degree 3 and two different types of lines

- the $2n$ old ordinary lines

- $n$ new dotted lines which indicate the pairing chosen at each vertex (see Figure 5).

\begin{figure}[!htb]
\centering
\includegraphics[scale=0.6]{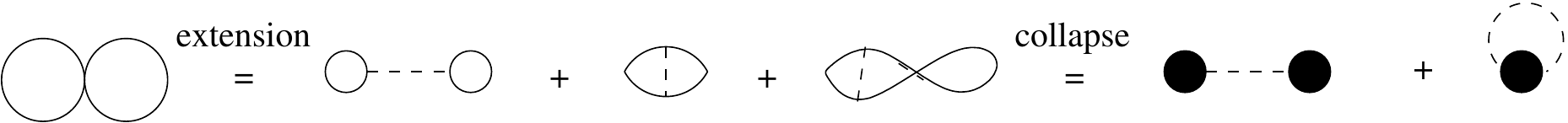}
\caption{The extension and collapse for the order 1 graph}
\label{odone}
\end{figure} 

Such graphs $G'$ are called the  3-body extensions of $G$ and we write
$G' {\ \rm ext\ } G$ when $G'$ is an extension of $G$. Let us introduce
for each such extension $G'$ an amplitude $A_{G'} = 3^{-n} A_G$ so that

\be  A_G =  \sum_{G' {\ \rm ext\ } G} A_{G'}
\ee
when $G'$ is an extension of $G$.

Now the ordinary lines of any extension $G'$ of any $G$
must form cycles. These cycles are joined by dotted lines. 

\begin{definition}
We define the collapse $\bar G'$ of such a $G'$ graph as
the graph obtained by contracting each cycle to a "bold" vertex (see Figure 5).
We write $\bar G' {\ \rm coll\ } G'$ if $\bar G'$ is the collapse of $G'$, and define
the amplitude of the collapsed graph $ \bar G'$ as equal to that of $G'$.
\end{definition}

Remark that collapsed graphs, made of bold vertices and dotted lines,
can have now arbitrary degree at each vertex.
Remark also that several different extensions of a graph $G$
can have different collapsed graphs, see Figure 5.

Now the loop vertex expansion rewrites
\be  S = \sum_G A_G  = \sum_{G' {\ \rm ext\ } G} A_{G'} =
 \sum_{\bar G' {\ \rm coll\ } G' {\ \rm ext\ } G}   A_{\bar G'} .
\ee 
Now we perform the tree repacking according to the graphs $\bar G'$
with the $n$ dotted lines and {\it not} with respect to $G$. This is a completely
different repacking:
\be  A_{\bar G'} = \sum_{\bar \cT \in \bar G'} w(\bar G', \bar \cT) A_{\bar G'},
\ee
so that
\be  S= \sum_{G' {\ \rm ext\ } G}   A_{\bar G'}  = \sum_{\bar \cT \in \bar G'} A_{\bar \cT},
\ee
\be
A_{\bar \cT} = \sum_{\bar G' \supset \bar \cT}  w(\bar G',\bar \cT) A_{\bar G'} .
\ee
The "miracle" is that 

\begin{theorem} For $\lambda$ small 
\be  \sum_{\bar \cT}  \vert A_{\bar \cT} \vert < \infty
\ee
the result being the {\it Borel sum} of the initial perturbative series \cite{R2}.
\end{theorem}

The proof of the theorem will not be recalled here 
(see  \cite{R1,MR1,R2}) but it relies on the positivity property of
the $ x^\cT_{\ell}(\{w\})$ symmetric matrix, and the representation of each
$A_{\bar \cT} $ amplitude as an integral over a corresponding normalized
Gaussian measure of a product of resolvents bounded by 1.  This convergence
would not be true if we had chosen naive $w(\cT,G)$ barycentric weights such as 1/5 for
each of the five trees of  the graph in Figure 1.

This method is valid for any $\phi^4$ model in any dimension with cutoffs \cite{MR1}.
It is not limited to $\phi^4$ but works eg for any stable interaction
at the cost of introducing more intermediate particles until three body 
elementary interactions are reached \cite{RW}. It also reproduces correctly
the large $N$ behaviour of $\phi^4$ matrix models, which was the key property for
which this expansion was found \cite{R1}.

\section{Examples}
In this section we give the extension and collapse of the Feynman graphs for $Z$ and $\log Z$ for the $\phi^4_0$ model up to order 3. 
We also recover the combinatorics of those graphs 
through the ordinary functional integral formula
for the loop vertex expansion formula of \cite{R2}.

\begin{figure}[!htb]
\centering
\includegraphics[scale=0.6]{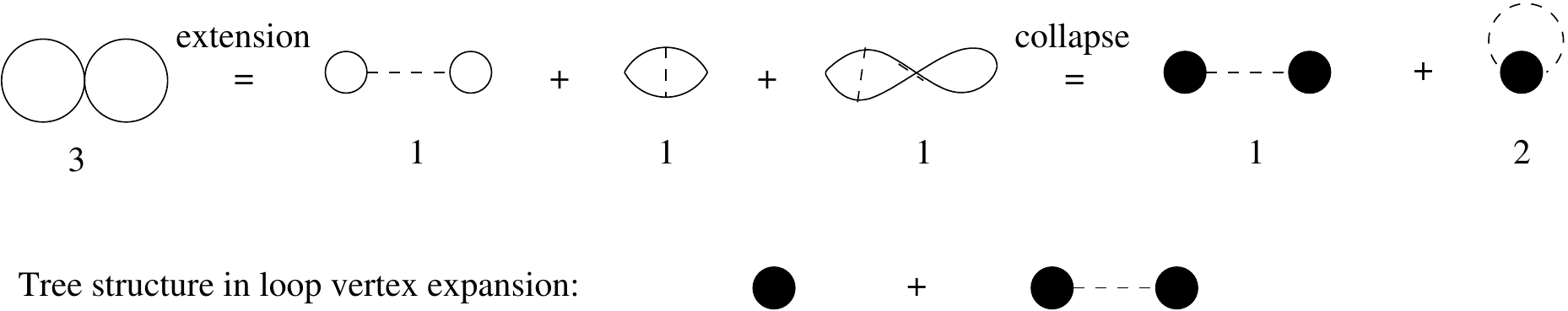}
\caption{The extension and collapse for order 1 graph, with combinatoric weight shown below, and the 
list of corresponding trees.}
\label{odonea}
\end{figure}

The extension and collapse at order 1 was shown in Figure (\ref{odonea}). In this case the tree structure is easy. We find only the trivial "empty" tree with one vertex and no edge and
the "almost trivial"  tree with two vertices and a single edge. The weight for these trees is 1.

At second order we find one disconnected Feynman graph and two connected ones. Only the
connected ones survive in the expansion of $\log Z$.
\begin{figure}[!htb]
\centering
\includegraphics[scale=0.6]{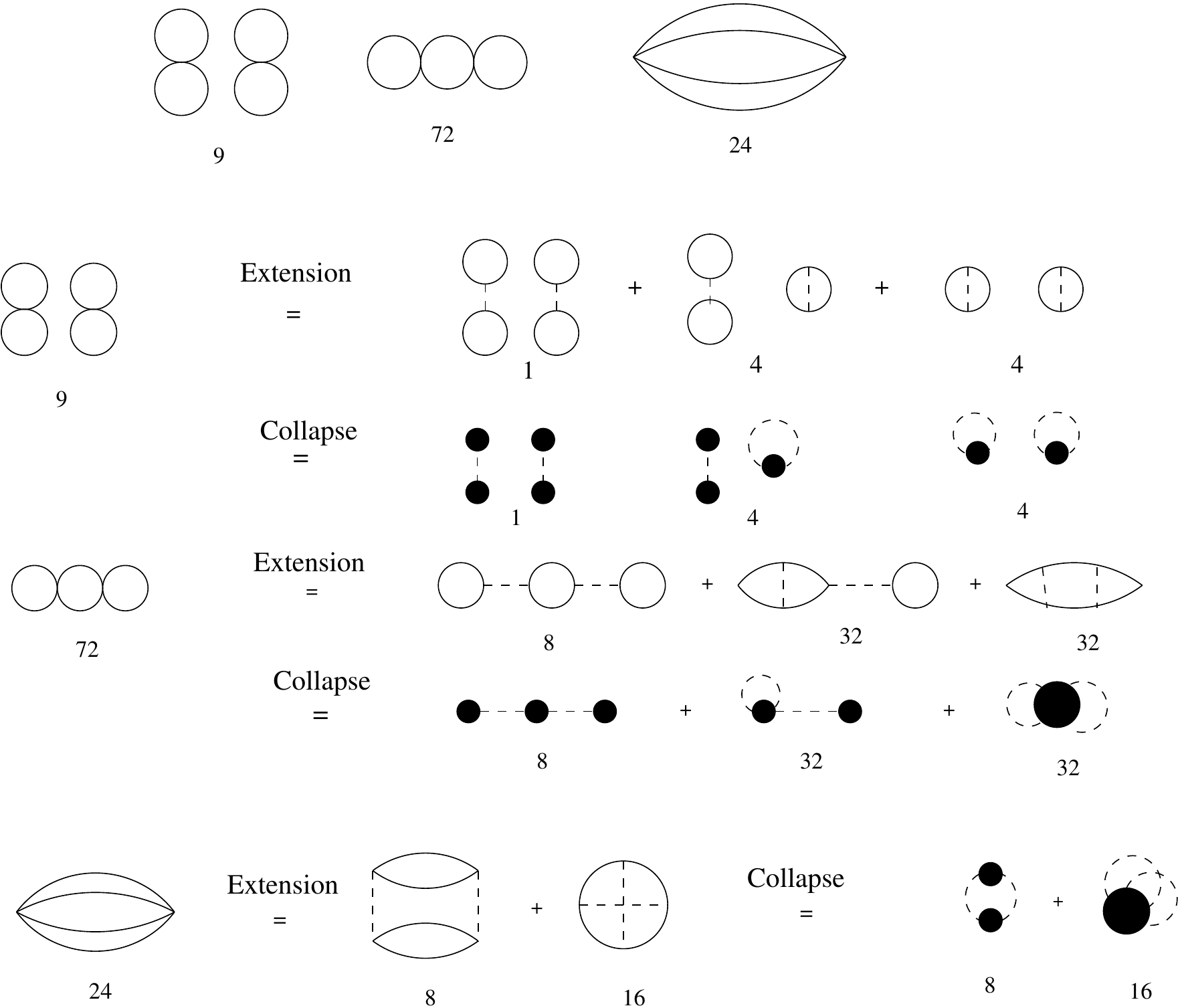}
\caption{The extension and collapse for order 2 graph and the number of graphs.}
\label{odtwo}
\end{figure} 

\begin{figure}[!htb]
\centering
\includegraphics[scale=0.7]{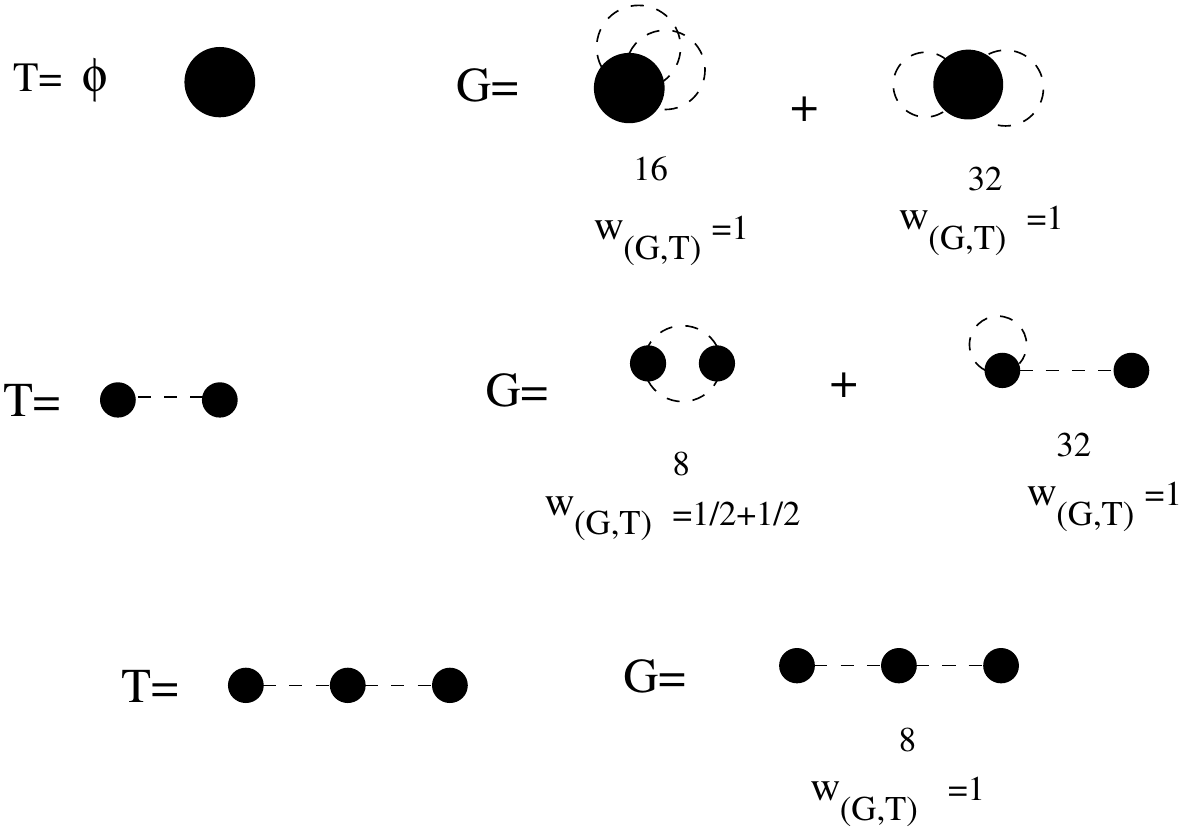}
\caption{The connected graphs and the tree structure from the Loop vertex expansion.}
\label{logtree}
\end{figure}

The corresponding graphs and tree structures are shown in Figure (\ref{odtwo}) and Figure(\ref{logtree}). Using the loop vertex expansion formula we begin to see that graphs that come from different order of the expansion of $\lambda$ are associated to the same trees by the loop vertex expansion. Indeed we recover contributions for the trivial and
almost trivial trees of the previous figure. But we find also a new contribution belonging to a tree with two edges.

\begin{figure}[!htb]
\centering
\includegraphics[scale=0.6]{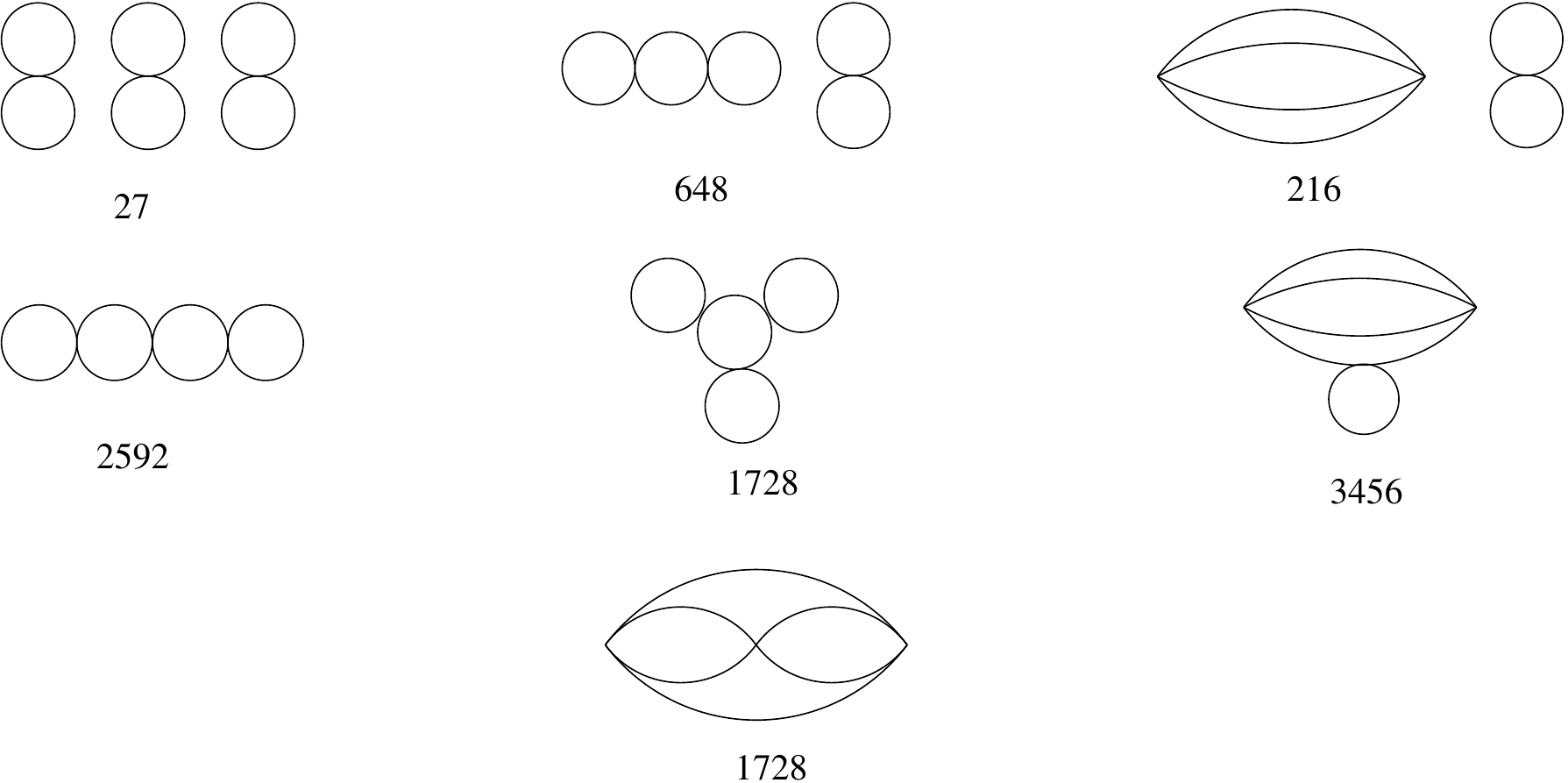}
\caption{The order 3 vacuum graph and the number of graphs.}
\label{odthree}
\end{figure} 

\begin{figure}[!htb]
\centering
\includegraphics[scale=0.53]{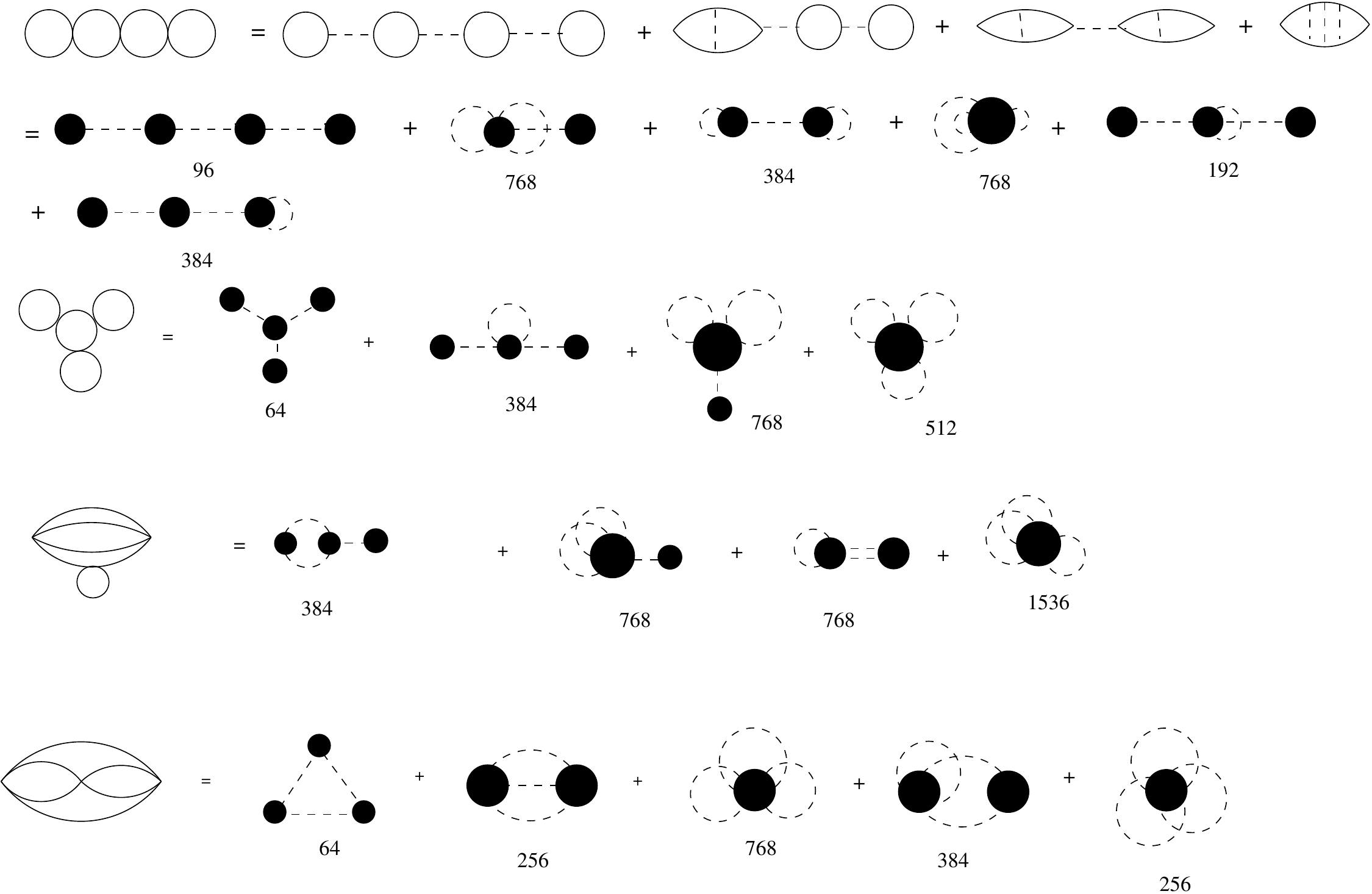}
\caption{The extension and collapse for order 3 graph.}
\label{ods}
\end{figure} 

\begin{figure}[!htb]
\centering
\includegraphics[scale=0.55]{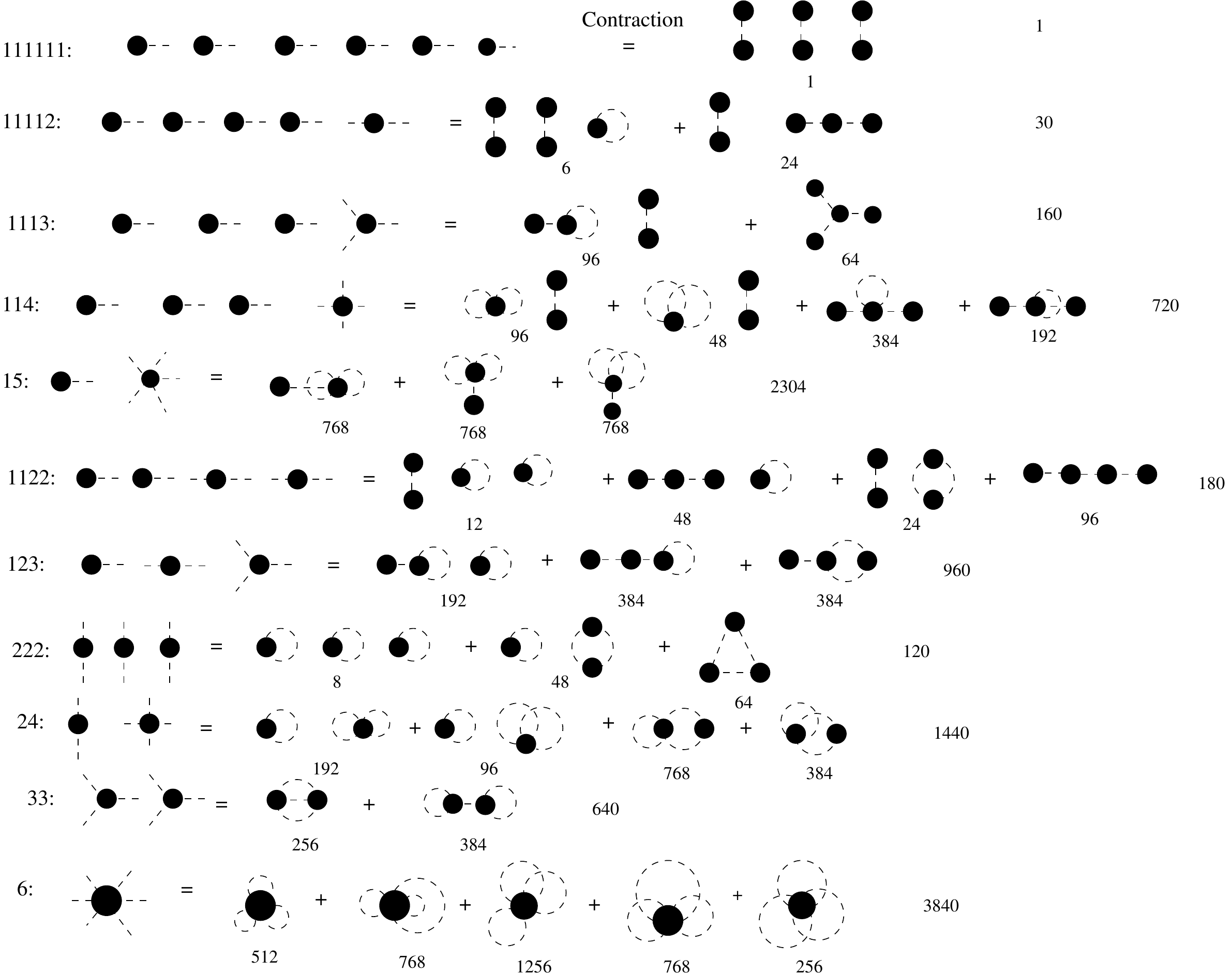}
\caption{The graph structure and combinatorics from the loop vertex expansion at order 3. The symbols like 1122 means we have 4 loop vertices V, two of them have one $\sigma$ field each and two of them have two $\sigma$ fields each, as we could read directly from this figure.}
\label{contraction}
\end{figure} 

\begin{figure}[!htb]
\centering
\includegraphics[scale=0.55]{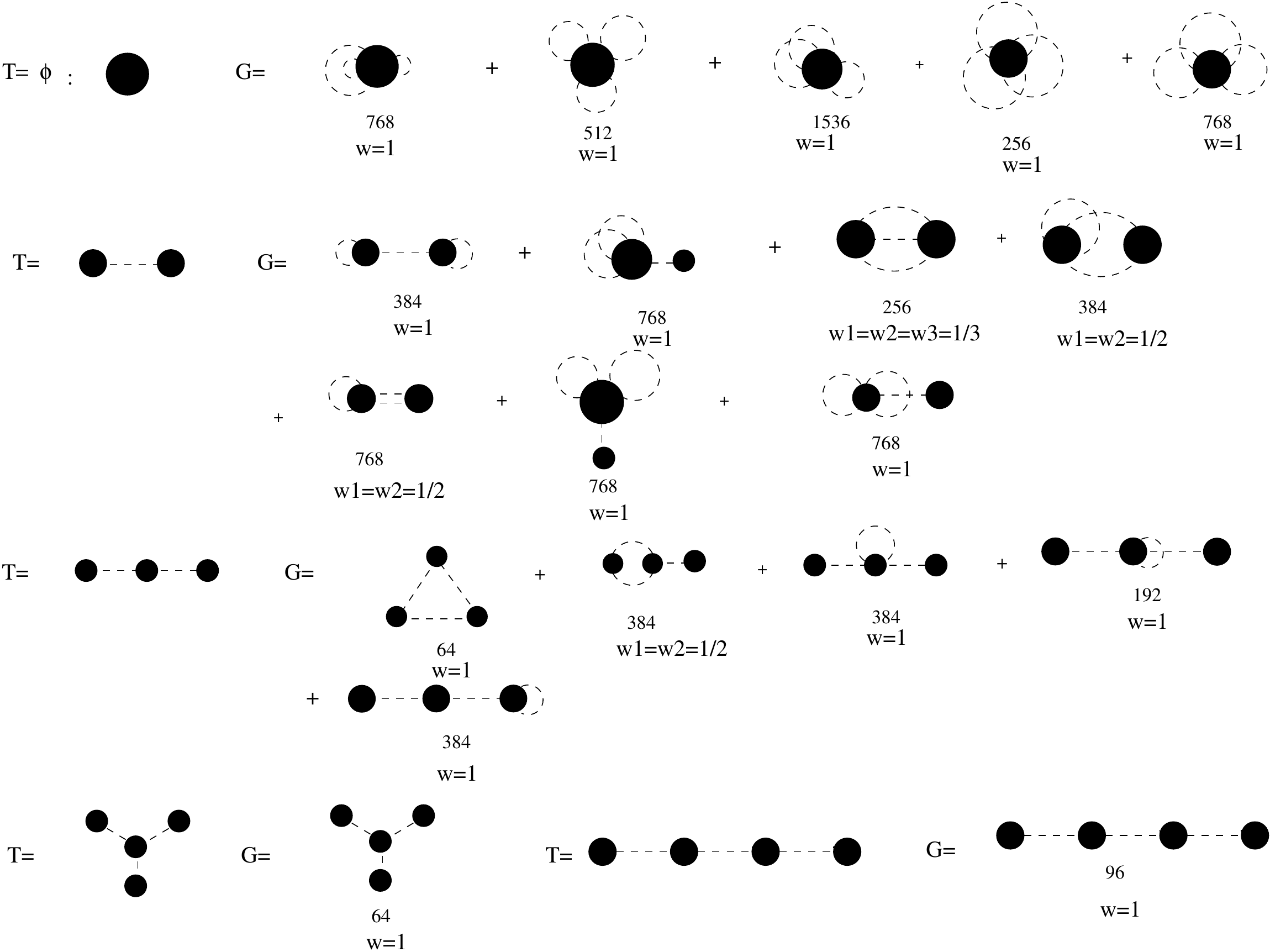}
\caption{The tree structure of order 3 graphs.}
\label{tree3}
\end{figure} 

At order three the computation becomes a bit more involved but the process is clear. We could start from the ordinary Feynman graphs and get the graphs of loop vertex expansion by extension and collapse. This is shown in Figure (\ref{ods}). The number under each collapsed graph means the number of the corresponding graphs, as in the previous case. The tree structure is shown in Figure(\ref{tree3}). In this figure the weight factor $w$ means always $w(G,\cT)$.
We could also get the graphs and combinatorics by using directly the loop vertex expansion, namely we integrate the $\phi$ fields and consider only the Wick contractions of the $\sigma$ fields. This is shown in the appendix and Figure ({\ref{contraction}}).  In this process we expand both $\exp V$ and the vertex $V=\tr\log(1+2i\sqrt{2\lambda}\sigma)$.  The interactions terms are then the loop vertices $V$ with various attached $\sigma$ fields. This is shown on the left hand of Figure ({\ref{contraction}}). For example, the symbol $123$ means we consider the $V^3/3!$ term in $\exp V$. We expand one of the $V$ to order $\lambda^{1/2}$, namely with one $\sigma$ field attached, one to the order $\lambda$, namely with 2 $\sigma$ fields attached and the third one to $\lambda^{3/2}$, namely with 3 $\sigma$ fields. Then we contract the sigma fields with respect to the Gaussian measure, obtaining all the contracted graphs. The total number of $123$ graphs could be read directly from this Gaussian integration. To get the 
combinatoric factor of each graph we need to compute the relative weights of these graphs. 
This is shown in the following example:
\begin{example}
\begin{figure}[!htb]
\centering
\includegraphics[scale=0.60]{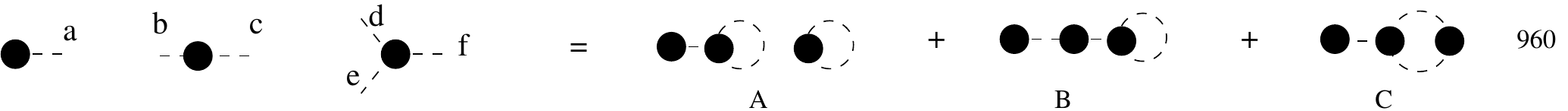}
\caption{The example of '123' contractions.}
\label{contracexample}
\end{figure} 
We consider the $123$ case for example. This is shown more explicitly in Figure(\ref{contracexample}). We use $a, b,\cdots, f$ to label the $\sigma$ fields attached to the  vertices. After the Wick contractions we get three different graphs $A$, $B$ and $C$. The number of possibilities to get $A$ is $3$, the number to get $B$ is $2\times 3=6$ and the number to get $C$ is also 6. So the relative weight for graph $A$ is $3/(3+6+6)=1/5$ and the relative weights for $B$ and $C$ are both $6/(3+6+6)=2/5$. As we could read directly from the loop vertex formula that the total number of $123$ contraction graphs is $960$, we get finally the combinatoric factor of graph $A$ to be $960\times 1/5=192$, and the corresponding factors for graphs $B$ and $C$ are $960\times 2/5=384$. This result agrees with the one coming from the Feynman graph computation.
\end{example}

From these examples we find that the structure of  loop vertex expansion is totally different from that of Feynman graph calculus. At each order of the loop vertex expansion we combine terms in different orders of $\lambda$.

\section{Non-integer Dimension}

Let us now consider, eg for $0< D \le 2$ the Feynman amplitudes for the $\phi^4_D$ 
theory. They are given by the following convergent parametric representation
\be  A_{D,G} = \int _0^\infty d\alpha \frac{e^{-m^2\sum_\ell \alpha_\ell}  }{U_G^{D/2}}
\label{KS}
\ee
where $m$ is the mass and $U_G$ is the Kirchoff-Symanzik polynomial for $G$
\be  U_G  = \sum_{\cT \in G}  \prod_{\ell \not \in \cT}  \alpha_\ell  .
\ee

All the previous decompositions working at the level of graphs, they are independent
of the space-time dimension. We can therefore repack the series of Feynman amplitudes
in non integer dimension to get the $D$ dimensional tree amplitude:
\be   A_{D,\bar \cT}  = \sum_{G \supset \bar \cT}  w(\bar \cT,G) A_{D,G}
\ee

We know that for $D=0$ and $D=1$ the loop vertex expansion is convergent.
Therefore it is tempting to conjecture , for instance at least for $D$ real and $0 \le D <2$ (that is when no ultraviolet divergences require renormalization) 

\begin{conjecture} For $\lambda$ small 
\be  \sum_{\bar \cT}  \vert A_{D,\bar \cT} \vert < \infty
\ee
the result being the Borel sum of the initial perturbative series.
\end{conjecture}

Progress on this conjecture would be extremely interesting as it would
allow to bridge quantum field theories in various dimensions of space time, and ie perhaps
justify the Wilson-Fisher $4-\epsilon$ expansion that allows good numerical approximate 
computations of critical indices in 3 dimensions.

We know however that when renormalization is needed, ie for $D \ge 2$,
this approach has to be completed with the introduction of the correct counterterms.
Presumably in this case the tree expansion should be adapted to select optimal
trees with respect to renormalization group scales. This is work in progress.

An other possible approach  to quantum field theory in non integer dimension, 
also based on the forest formula but  more radical,
is proposed in \cite{GMR}.

\section{Conclusion}

The lessons we may draw from the Loop Vertex Expansion are

\begin{itemize}

\item Interactions should be decomposed into three body elementary interactions.
The corresponding fields might be more fundamental than the initial ones.

\item Tree formulas solve the constructive problem ie resum perturbation theory
at the cost of loosing locality of the new vertices. 

\end{itemize}

It may be also interesting to further understand why trees are 
so central both in the parametric formulas (\ref{KS})
for {\it single} Feynman amplitudes and in the non-perturbative treatment of the theory.
The answer might imply a complete refoundation of quantum field
theory around the notion of trees, rather than Feynman graphs or even
functional integrals \cite{GMR}.

\section{Appendix}

In this Appendix we compute the weight of collapsed Feynman graphs using the
Loop Vertex Expansion.

For the $\phi^4_0$ model we have:
\begin{equation}
 Z=\frac{1}{\sqrt{2\pi}}\int d\phi e^{-\frac{1}{2}\phi^2-\lambda\phi^4}=\frac{1}{\sqrt{2\pi}}\int d\sigma e^{-\frac{1}{2}\sigma^2-
\frac{1}{2} \log(1+2i\sqrt{2\lambda}\sigma)} .
\end{equation}
We define
\begin{equation}
 V=\frac{1}{2} \log (1+2i\sqrt{2\lambda}\sigma).
\end{equation}
In what follows we compute the vacuum graphs up to order $3$ in $\lambda$.
We expand $Z$ into powers of $V$:
\begin{equation}
Z=\frac{1}{\sqrt{2\pi}}\int d\sigma e^{-\frac{1}{2}\sigma^2} [1 - V+\frac{1}{2!}V^2 - \frac{1}{3!}V^3+\frac{1}{4!}V^4 - \frac{1}{5!}V^5+\frac{1}{6!}V^6] ,
\end{equation}
and we have
\begin{eqnarray}
\log(1+2i\sqrt{2\lambda}\sigma)&=&2\sqrt{2\lambda}i\sigma+4\lambda\sigma^2-\frac{16\sqrt{2}i}{3}\lambda^{3/2}\sigma^3-16\lambda^2
\sigma^4\nonumber\\
&+&\frac{128\sqrt{2}i}{5}\lambda^{5/2}\sigma^5+\frac{256}{3}\lambda^3\sigma^6 .
\end{eqnarray}

The first term
\begin{equation}
 \frac{1}{\sqrt{2\pi}}\int d\sigma e^{-\frac{1}{2}\sigma^2} 1=1
\end{equation}
is trivial.

The order $V$ terms give:
\begin{eqnarray}
&&-\frac{1}{\sqrt{2\pi}}\int d\sigma e^{-\frac{1}{2}\sigma^2}V=\frac{1}{\sqrt{2\pi}}\int d\sigma e^{-\frac{1}{2}\sigma^2}[-2\lambda\sigma^2+8\lambda^2\sigma^4-\frac{128}{3}\lambda^3\sigma^6]\nonumber\\
&=&-2\lambda+24\lambda^2-640\lambda^3 .
\end{eqnarray}
The $V^2$  terms give:
\begin{eqnarray}
&&\frac{1}{2!}\frac{1}{\sqrt{2\pi}}\int d\sigma e^{-\frac{1}{2}\sigma^2}V^2=\frac{1}{2!}(\frac{1}{2})^2\frac{1}{\sqrt{2\pi}}\int d\sigma e^{-\frac{1}{2}\sigma^2}[-8\lambda\sigma^2+16\lambda^2\sigma^4\nonumber\\
&-&\frac{64\times8}{9}\lambda^3\sigma^6+\frac{128}{3}\lambda^2\sigma^4-\frac{128\times 8}{5}\lambda^3\sigma^6-128\lambda^3\sigma^6]\nonumber\\
&=&-\lambda+22\lambda^2-\frac{320}{3}\lambda^3-624\lambda^3 .
\end{eqnarray}
The $V^3$ terms give:
\begin{eqnarray}
&&-\frac{1}{3!}\frac{1}{\sqrt{2\pi}}\int d\sigma e^{-\frac{1}{2}\sigma^2}V^3=-\frac{1}{3!}(\frac{1}{2})^3\frac{1}{\sqrt{2\pi}}\int d\sigma e^{-\frac{1}{2}\sigma^2}[64\lambda^3\sigma^6-96\lambda^2\sigma^4\nonumber\\
&+&384\lambda^3\sigma^6+512\lambda^3\sigma^6]\nonumber\\
&=&6\lambda^2-300\lambda^3 .
\end{eqnarray}
The $V^4$ terms give:
\begin{eqnarray}
&&\frac{1}{4!}(\frac{1}{2})^4\frac{1}{\sqrt{2\pi}}\int d\sigma e^{-\frac{1}{2}\sigma^2}[64\lambda^2\sigma^4-
\frac{2048}{3}\lambda^3\sigma^6-768\lambda^3\sigma^6]\nonumber\\
&=&\frac{1}{2}\lambda^2-\frac{80}{3}\lambda^3-30\lambda^3 .
\end{eqnarray}
The $V^5$ terms give:
\begin{eqnarray}
-\frac{1}{5!}(\frac{1}{2})^5\frac{1}{\sqrt{2\pi}}\int d\sigma e^{-\frac{1}{2}\sigma^2}\ 1280\lambda^3\sigma^6=-5\lambda^3 .
\end{eqnarray}
The $V^6$ term gives:
\begin{eqnarray}
-\frac{1}{6!}(\frac{1}{2})^6\frac{1}{\sqrt{2\pi}}\int d\sigma e^{-\frac{1}{2}\sigma^2}\ 512\lambda^3\sigma^6=-\frac{1}{6}\lambda^3 .
\end{eqnarray}
So up to $3$rd order in $\lambda$ we recover
\begin{equation}
 Z=-3\lambda+\frac{105}{2}\lambda^2-\frac{10395}{6}\lambda^3=-4!! \lambda+\frac{8!!}{2!}\lambda^2-\frac{12!!}{3!}\lambda^3 ,
\end{equation}
which of course coincide with the 
number of ordinary Wick contractions derived 
by the regular $\lambda\phi^4$ Feynman expansion.

\medskip
\noindent{\bf Acknowledgments}
We thank H.  Kn\"orrer for asking the question which lead to writing this paper.

\end{document}